# Selective hydrogenation improves interface properties of high-*k* dielectrics on 2D semiconductors

Yulin Yang[1, #], Tong Yang[2, #], Tingting Song[3], Jun Zhou[4], Jianwei Chai[4], Lai Mun Wong[4], Hongyi Zhang[1], Wenzhang Zhu[1], Shijie Wang[4, *], and Ming Yang[2, *]

[1] Fujian Provincial Key Laboratory of Optoelectronic Technology and Devices, School of Optoelectronic and Communication Engineering, Xiamen University of Technology, Xiamen, 361024, China

[2] Department of Applied Physics, The Hong Kong Polytechnic University, Hung Hom, Kowloon, Hong Kong SAR, China

[3] College of Physics and Space Science, China West Normal University, Nanchong, 637002, China

[4] Institute of Materials Research and Engineering, Agency for Science, Technology and Research (A*STAR), 2 Fusionopolis Way, Innovis, 138634, Singapore

[#] These authors contributed equally to this work.

[*] Author to whom correspondence should be addressed:

M.Y.(mingyang@polyu.edu.hk) or S.J.W. (sj-wang@imre.a-star.edu.sg)

**ABSTRACT**:

The integration of high-*k* dielectrics with two-dimensional (2D) semiconductors is a critical step towards high-performance nanoelectronics, which however remains challenging due to high density of interface states and the damage to the monolayer 2D semiconductors. In this study, we propose a selective hydrogenation strategy to improve the interface properties while do not affect the 2D semiconductors. Using the interface of monolayer MoS$_2$ and silicon nitride as an example, we show substantially improved interface properties for electronic applications after the interfacial hydrogenation, as evidenced by reduced inhomogeneous charge redistribution, increased band offset, and untouched electronic properties of MoS$_2$. Interestingly, this hydrogenation process selectively occurs only at the silicon nitride surface and is compatible with the current semiconductor fabrication process. We further show that this strategy is general and applicable to other interfaces between high-*k* dielectrics and 2D semiconductors such as HfO$_2$ on the monolayer MoS$_2$. Our results demonstrate a *simple yet viable* way to improve the interfacial properties for integrating many high-*k* dielectrics on a broad range of two-dimensional transition metal disulfide semiconductors.

**INTRODUCTION**

Two-dimensional (2D) semiconducting materials are appealing for nanoelectronic applications in the post-Moore era as their ultra-thin thickness can help minimize the short channel effect confronted with current Si based technologies.[1, 2] Among them, monolayer transition metal disulfides (TMDs) such as $MoS_2$ and $WS_2$ have attracted tremendous interest as promising channel materials for electronic or optoelectronic devices due to their stability, direct band gaps, tunable electronic and optical properties, and many other excellent physical properties.[3-10] The monolayer $MoS_2$ based nanotransistors have shown encouraging device performance such as a high on/off ratio.[11-13] However, to utilize monolayer TMDs in practical electronic applications, there remain many challenges[14], which include the large-scale growth of high-quality TMDs layers [15-19], metal contact with low resistance[20-25], and the integration of high-$k$ dielectrics.[26-29]

In particular, for electronic applications, the integration of high-$k$ dielectrics is highly desired as their large dielectric constants can effectively screen the scattering from charge impurities and thus improve the device's performance.[30, 31] However, it is found difficult to achieve high-quality interfaces between high-$k$ dielectrics and monolayer TMDs which results in high interface state density.[28, 29] Currently, a high-quality interface can be realized by van der Waals (vdW) integration of $h$-BN or $CaF_2$ layers on TMDs semiconductors[32-34], but the sophisticated integration process and the difficulty in direct growth of high-quality vdW layers impede its large-scale practical applications. While various attempts have been made to directly deposit high-$k$ oxides such as $HfO_2$[35], $Al_2O_3$[36, 37] and $ZrO_2$[38] on TMDs layers using atomic layer deposition (ALD) or sputtering technique, their interfacial properties are still much inferior to that of $HfO_2$/Si and need further improvement.[29, 39]

One such endeavor is interface engineering, which includes interface passivation by annealing [40, 41], and utilizing interface seeding or buffer layers[28, 29, 42]. These strategies have brought some improvement from their respective effects but may also raise new problems such as the damage of the 2D TMDs semiconductors and accordingly deteriorated electronic properties. [28, 29] Thus, it is highly desirable to develop a holistic strategy that can effectively passivate the interfacial dangling bonds of high-*k* dielectrics but does not deteriorate the 2D semiconductors. In this study, we propose selective hydrogenation as such an ideal strategy, thanks to the fact that the hydrogen atoms are prone to be adsorbed on the surface of high-*k* dielectrics but inert to the basal planes of 2D TMDs. We use the interface between monolayer $MoS_2$ and *β*-$Si_3N_4$ (0001) as an example to deliver the concept and apply this strategy to another interface between $MoS_2$ and $HfO_2$ to demonstrate its generality.

**METHOD**

All the calculations were performed using density-functional theory (DFT) based the Vienna Ab initio Simulation Package (VASP.5.4.4.18) with the Perdew-Burke-Ernzerhof (PBE) functional and the projector-augmented wave (PAW) potentials.[43-45] The cutoff energy for the plane wave expansion was set to 500 eV. Γ-centered 9×9×4, 12×12×1, 6×6×1, and 9×9×1 *k*-point meshes were used to sample the first Brillouin zone of bulk *β*-$Si_3N_4$, monolayer $MoS_2$, and the interface of monolayer $MoS_2$ on *β*-$Si_3N_4$ (0001) and *c*-$HfO_2$ (111), respectively. A vacuum layer with a thickness of 15 Å was used for all slab structures to minimize the artificial Coulomb interaction between two adjacent surfaces. For the interface structures, the vdW correction has been included by Grimme's DFT-D3 method.[46] For all calculations, the electronic and ionic convergence criteria were set to $10^{-5}$ eV and 0.01 eV/Å, respectively. The calculated lattice constants of *β*-$Si_3N_4$ and $MoS_2$ monolayer are

a=b=7.659 Å and c=2.925 Å, and 3.184 Å, and the corresponding PBE band gaps are 4.24 eV and 1.66 eV, respectively. All these results are well consistent with previous studies.[47-51]

The interface structures were constructed by placing the (4×4×1) monolayer MoS$_2$ supercell on the ($\sqrt{3} \times \sqrt{3} \times 1$) β-Si$_3$N$_4$ (0001) surface without/with surface hydrogen passivation, in which a 3.97% compressive strain was applied to the Si$_3$N$_4$. This strain yields a reduced PBE band gap by ~0.2 eV in the Si$_3$N$_4$ (see Fig. S1 in the Supporting Information), while the main electronic structure does not change much compared with that of the pristine bulk Si$_3$N$_4$. The thickness of the β-Si$_3$N$_4$ (0001) surface was set to 7 atomic layers, the bottom layer of which was passivated using hydrogen atoms. In the interface structures, dipole correction was applied.[52] To examine interfacial stability of monolayer MoS$_2$/hydrogen passivated β-Si$_3$N$_4$ (0001), *ab initio* molecular dynamics simulations were performed on the interface supercell, where the canonical ensemble (NVT) and the Nosé heat bath were adopted with a time step of 1 fs and a time length of 6 ps

The interfacial interaction strength between monolayer MoS$_2$ and dielectrics (D, Si$_3$N$_4$ or HfO$_2$) can be estimated from the adsorption energy ($E_{ad}$) as defined below:

$$E_{ad} = E_{\text{MoS}_2+D} - E_{\text{MoS}_2} - E_D, \qquad (1)$$

where $E_{\text{MoS}_2+D}$ is the total energy of the hybrid interface structure for MoS$_2$ on the dielectrics, and $E_{MoS_2}$ and $E_D$ are the total energy of the isolated MoS$_2$ monolayer and the dielectric (Si$_3$N$_4$ or HfO$_2$) surface, respectively. Similarly, the charge redistribution Δρ for MoS$_2$ monolayer on the dielectric surface is defined as:

$$\Delta\rho = \rho_{\text{MoS}_2+D} - \rho_{\text{MoS}_2} - \rho_D, \qquad (2)$$

in which $\rho_{\text{MoS}_2+D}$ is the charge density of the hybrid interface structure for MoS$_2$ on the dielectric, and $\rho_{\text{MoS}_2}$ and $\rho_D$ are the charge density of the isolated MoS$_2$ monolayer and the dielectric (Si$_3$N$_4$ or HfO$_2$) surface, respectively.

**RESULTS AND DISCUSSIONS**

Silicon nitride (Si$_3$N$_4$) has a large band gap and high thermal stability, which has been widely applied in current Si based electronic devices.[49, 50, 53] It was reported that high-quality Si$_3$N$_4$ thin films can be deposited on graphene[48, 54, 55], while the electronic properties of the latter does not change much.[54] Furthermore, Si$_3$N$_4$ might be a better choice than the metal oxides as a high-*k* dielectric because it does not contain metallic ions which tend to interact strongly with 2D TMDs.[47, 56, 57] Thus, in this study, we use the interface between monolayer MoS$_2$ and *β*-Si$_3$N$_4$ as a model example to study the hydrogenation effect on the interface properties.

We have constructed various interface structures by sliding the MoS$_2$ on the Si$_3$N$_4$ surface. The energy difference among these configurations is within 12 meV (see Fig. S2(a)), inferring a weak interfacial interaction. The top and side view of the most stable interface configuration is shown in Fig. S2(b) and Fig 1(a), respectively, in which the interfacial S atoms of MoS$_2$ tend to arrange near the interfacial Si atoms of Si$_3$N$_4$ to maximize the potential bonding. Similar trend has been reported at the interfaces of MoS$_2$/HfO$_2$ or high-*k* dielectrics/graphene.[47, 58, 59] The interface spacing between monolayer MoS$_2$ and *β*-Si$_3$N$_4$ (0001) is about 2.9 Å, at the lower bound of the vdW interaction range, further indicating weak interface interaction. The interface interaction strength can also be measured from the adsorption energy (-23.2 meV/Å$^2$). This value is slightly higher than that of bi-layer MoS$_2$ (-22.4 meV/Å$^2$), but it is much smaller than those of other high-k dielectric/TMDs interfaces such as $MoS_2/HfO_2$ and $MoS_2/SiO_2$.[47, 57, 60]

The projected density of states (PDOSs) of monolayer $MoS_2$ on β-$Si_3N_4$ (0001) is shown in Figs. 1(c-e). One important observation is that the electronic structure of monolayer $MoS_2$ is nearly intact in the presence of the $Si_3N_4$ due to relatively weak interaction between them. As the PDOS shown in Fig. 1(c), the monolayer $MoS_2$ remains semiconducting with a PBE gap of ~1.6 eV. This is highly desired because a sizable band gap in semiconductors enables a stable device operation with a large on/off ratio. However, due to the dangling bonds in β-$Si_3N_4$ (0001) (see Fig. S3) induced mid-gap states, orbital hybridization is seen between $MoS_2$ and $Si_3N_4$. As Figs. 1(b-d) show, the Mo $e_g$ orbitals couple with the p orbitals of Si and N ions near the valence band edge, while at the conduction edge, the hybridization among Mo $e_g$, S $p_z$ orbital and Si $p_z$ orbital is more noticeable. This orbital hybridization leads to insignificant band offsets between $MoS_2$ and $Si_3N_4$, which is unfavorable for device applications. To minimize the tunneling current in semiconductor devices, the valence and conduction band offset between the semiconductor (e.g., $MoS_2$ in this study) and the gate dielectric (e.g., $Si_3N_4$ here) should be larger than 1 eV.[30] In addition, the interfacial interaction causes charge redistribution at the interface (see Fig. 1(a)), where the accumulated charges are closer to $Si_3N_4$ surface while the depleted charges are more pronounced near $MoS_2$ side. The charge redistribution induces electron-hole puddles in monolayer $MoS_2$, as shown in Fig. 1(b). This is another detrimental effect for electronic device applications as the electron-hole puddles can be charge scattering centers which reduce the carrier mobility.[61]

Since the degraded interface properties are mainly due to the dangling bonds at the β-$Si_3N_4$ (0001) surface, one natural attempt for improving the interface properties is to passivate them. We note that the hydrogenation process has been widely used in semiconductor technologies to passivate the intrinsic defects.[62] Such process may also be applied to passivate the dangling bonds of the $\beta$-$Si_3N_4$ (0001) surface considered in this study. After passivation (see the atomic structures in Fig. S4(a)), the dangling bonds induced states nearly vanish as confirmed by the PDOSs in Fig. S4(b). Consequently, the interfacial interaction between monolayer $MoS_2$ and hydrogen passivated $\beta$-$Si_3N_4$ (0001) surface is further weakened. The calculated adsorption energy (see Fig. 2(a)) decreases slightly to -18.2 meV/Å$^2$. The charge redistribution is shown in Fig. 2(b) and (c), which clearly suggests that the weaker interfacial interaction results in much less pronounced interfacial charge redistribution, as well as the suppressed electro-hole puddles on the Mo atomic plane.

After the hydrogen passivation on the $\beta$-$Si_3N_4$ (0001) surface, the orbitals near Fermi level between monolayer $MoS_2$ and $Si_3N_4$ are well separated. From Fig. 2(d), we can see that the valence band edge now is contributed by p orbital from $Si_3N_4$, while the states contributed by Mo d orbital are located at -0.98 eV below the Fermi level. In contrast, the conduction band edge is mainly derived from Mo d orbital, and the contribution from p orbital of $Si_3N_4$ starts from ~4.88 eV and above. This leads to a type-II band offset between monolayer $MoS_2$ and $Si_3N_4$, in which the valence band (VBO) and conduction band offset (CBO) can be estimated to be 0.98 and 3.3 eV at the PBE level, respectively. It should be noted that these band offset values might be underestimated, which could be larger if the quasiparticle corrections or hybrid functionals are used. With the hydrogen passivation, we also find improvement on the electronic properties of monolayer $MoS_2$. As shown in Fig. 2(e), the electronic structure of monolayer $MoS_2$ on the hydrogen passivated $Si_3N_4$ is nearly identical to that of free-standing monolayer $MoS_2$. Thus, we can see that using this simple hydrogenation process, the interface properties of MoS$_2$/Si$_3$N$_4$ are improved remarkably as evidenced by the large band offsets, suppressed electron-hole puddles, and intact electronic properties of MoS$_2$.

More importantly, the hydrogenation process is energetically selective to *β*-Si$_3$N$_4$ (0001) surface. As shown in Fig. 3(a) shows, the adsorption energy for hydrogen atom on top of Si site at the Si$_3$N$_4$ surface is -0.23 eV, and the lowest adsorption energy on the top of the N sites is -1.11 eV due to the dangling bonds of N ions at the surface. The negative adsorption energy indicates that the hydrogen adsorption is energetically favorable, and the hydrogenation can occur spontaneously even at low temperature. In contrast, the energy for hydrogen atoms adsorbed on the monolayer MoS$_2$ is calculated to be 1.84 eV, consistent with previous studies.[63] The large positive adsorption energy suggests that the hydrogenation on the monolayer MoS$_2$ is energetically unfavorable, which is less likely to happen even at a high processing temperature. Further quantum mechanical molecular dynamics simulation shows that the hydrogen passivated MoS$_2$/Si$_3$N$_4$ interface is thermodynamically stable at the high temperature of 800 K. From Fig. 3(b) and Fig. S(5), it is noted that the variation of interfacial Si-H and N-H bond length is within 0.2 Å during the MD simulation, suggesting that the Si$_3$N$_4$ surface adsorbs hydrogen atoms so strongly that they are unlikely to diffuse to monolayer MoS$_2$ even at a temperature of 800 K (see Fig. 3(c)). This implies that the spontaneously selective hydrogenation process is compatible and stable with current semiconductor device fabrication processes, as most of them are conducted below 800 K.[64]

Next, we show that the hydrogenation process is applicable for improving interface properties of other high-*k* oxides and monolayer MoS$_2$. Since HfO$_2$ has been widely used in current electronic devices, we use it as a model example to further examine the hydrogenation effect. It has been reported that Hf-terminated $HfO_2$ interacts strongly with $MoS_2$, leading to inferior electronic properties, while the interface between O-terminated $HfO_2$ and $MoS_2$ shows improved interface performance.[47] Thus, we focus on the interface of Hf-terminated $HfO_2$ and monolayer $MoS_2$ without/with the interfacial hydrogenation. As Figs. 4(a) and (c) show, at the interface of Hf-terminated $HfO_2$ and monolayer $MoS_2$, it forms interfacial covalent Hf-S bonds. This induces noticeable charge transfer from $HfO_2$ into $MoS_2$, making the $MoS_2$ metallic. Similarly, the interface properties can be improved using the hydrogenation process. Figures 4(b) and (d) are the charge redistribution and local density of states (LDOS) for monolayer $MoS_2$ on the hydrogen passivated Hf-$HfO_2$ (0001) surface, respectively. The reduced interfacial charge redistribution and the suppressed mid-gap metallic states in $MoS_2$ clearly suggest that better interface properties are reached by the selective hydrogen passivation.

It has been noted that 2D materials with $HfO_2$ thin films grown by atomic layer deposition (ALD) technique show better device performance than those using sputtering or physical vapor deposition (PVD).[65] We argue that the underlying mechanisms, in addition to the film uniformity and quality, could be partially ascribed to unintended hydrogen passivation on the high-*k* oxide film. In ALD process, the Hf metal-based precursor and water vapor are used for the deposition of $HfO_2$[66], in which residual hydrogen source might evolve to passivate the dangling bonds, whereas the high vacuum required in sputtering or PVD process leads to negligible residual hydrogen sources. Thus, we believe that the interface between high-*k* dielectrics and $MoS_2$ can be further improved by intentionally depositing the high-*k* dielectric films in the hydrogen environment at the first few cycles. Please note that the effectiveness of this interfacial hydrogenation strategy is strongly dependent on the difference in the hydrogen adsorption on 2D materials and high-$k$ dielectrics. If hydrogen interacts strongly on the 2D materials such as graphene[67] or phosphorene[68], the hydrogenation process would result in undesired hydrogen adsorption on the 2D materials as it will change the electronic properties of 2D materials. On the contrary, as long as a 2D material is unfavorable to bind hydrogen, the hydrogenation process proposed here is beneficial to its integration with high-$k$ dielectrics. The beneficial effects of this hydrogenation strategy on the interface integration are two-fold: one is to passivate the dangling bonds in the high-$k$ dielectrics, resulting in reduced interface state density; the other is to passivate the native defects in 2D TMDs, which can further improve electronic properties of the 2D semiconductors.

In conclusion, we report a simple yet effective strategy to improve the interfacial properties of high-k dielectrics and 2D semiconductors, as supported by the substantially improved electronic properties at the interface between $Si_3N_4$ and monolayer $MoS_2$ using the proposed selective hydrogenation process. This hydrogenation process spontaneously occurs on $Si_3N_4$, leaving the monolayer $MoS_2$ nearly intact. We reveal that the hydrogenated $MoS_2/Si_3N_4$ interface is stable at high temperatures, which is compatible with the current semiconductor fabrication process. We further show that the interfacial hydrogenation strategy can be applied for the integration of other high-$k$ dielectrics on $MoS_2$ monolayer. Given similar surface chemistry among TMD semiconductors, this hydrogenation process can be extended to the interfaces of many high-$k$ dielectrics and a broad range of TMDs such as $MoS_2$, $WS_2$, or $HfS_2$, as well as some transition metal dichalcogenides such as $MoSe_2$ or $WSe_2$, or other 2D materials that are inert to hydrogen adsorption, enabling us to boost the development of 2D semiconductors based nanoelectronic devices.

**Acknowledgements**

M.Y. acknowledges the funding support (project IDs: 1-BE47 and ZE2F) from The Hong Kong Polytechnic University. We acknowledge Centre for Advanced 2D Materials and Graphene Research at National University of Singapore, and the National Supercomputing Centre of Singapore for providing computing resources.

**Conflict of Interest**

The authors declare no conflict of interest.

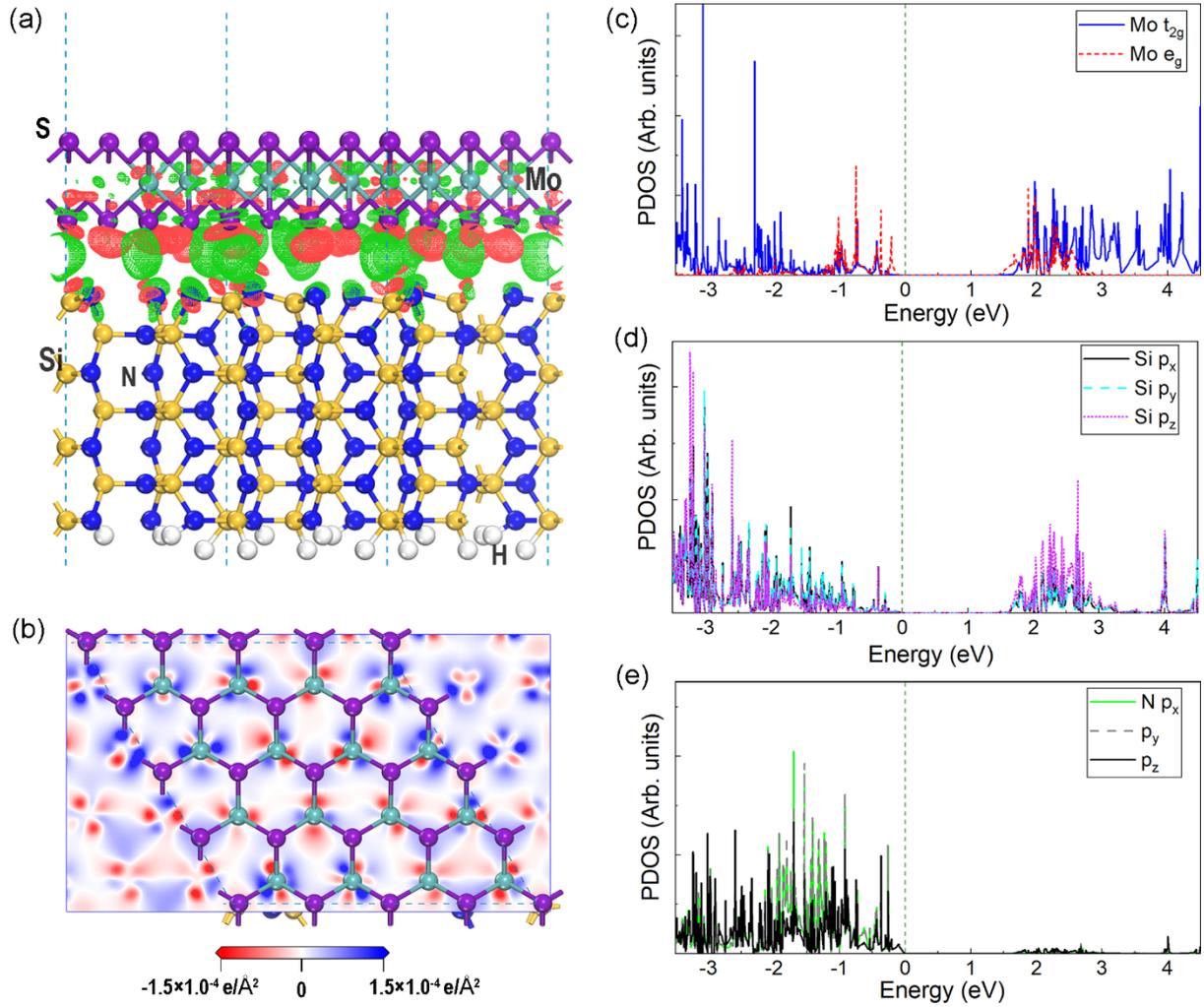

**Figure 1.** Interface properties between monolayer MoS$_2$ and *β*-Si$_3$N$_4$ (0001). (a) top and (b) side view of the most stable configuration for monolayer MoS$_2$ on *β*-Si$_3$N$_4$ (0001), in which the red and green dots in (a) denote the depleted and accumulated charge density visualized by an iso-surface value of 1.5×10$^{-4}$ e/Å$^3$, respectively. The red and blue colour in (b) denotes charge puddles on the Mo plane. The projected density of states (PDOSs) for Mo (c), Si (d) and N (e), where the Fermi level is set to 0 eV.

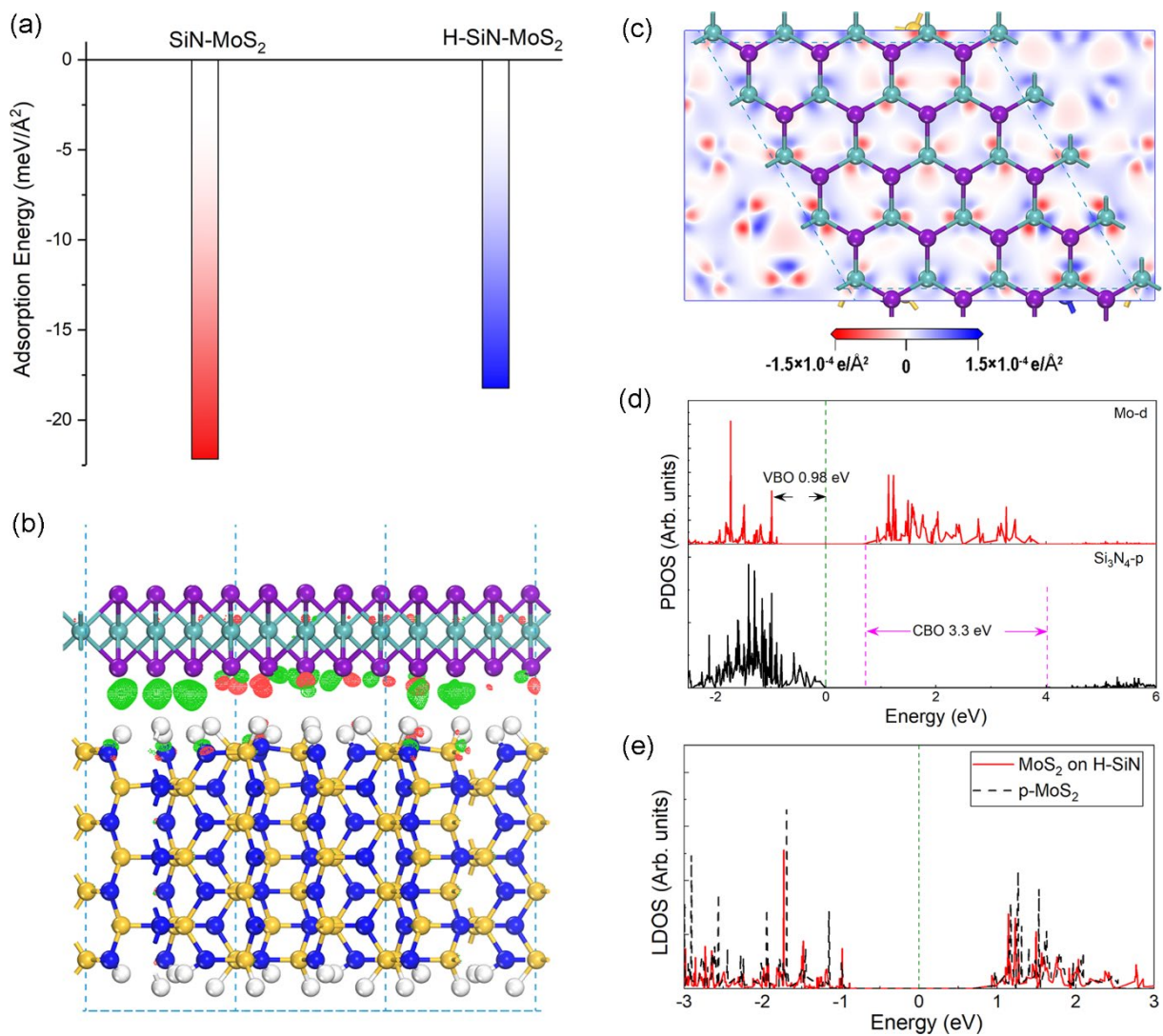

**Figure 2.** Interface properties of monolayer MoS$_2$ on hydrogenated *β*-Si$_3$N$_4$ (0001). (a) Adsorption energy of monolayer MoS$_2$ on *β*-Si$_3$N$_4$ (0001) w/o surface hydrogenation. (b) The side view of interface structure for monolayer MoS$_2$ on hydrogenated *β*-Si$_3$N$_4$ (0001) superimposed with visualized charge density redistribution using an iso-surface value of 1.5×10$^{-4}$ e/Å$^3$, in which red and green dots denote the depleted and accumulated charge density, respectively. (c) The charge puddles distributed on the Mo plane. (d) The PDOSs on Mo d orbital and p orbitals of central Si$_3$N$_4$ layer. (e) LDOSs of MoS$_2$ monolayer on hydrogenated *β*-Si$_3$N$_4$ (0001) and the pristine monolayer MoS$_2$.

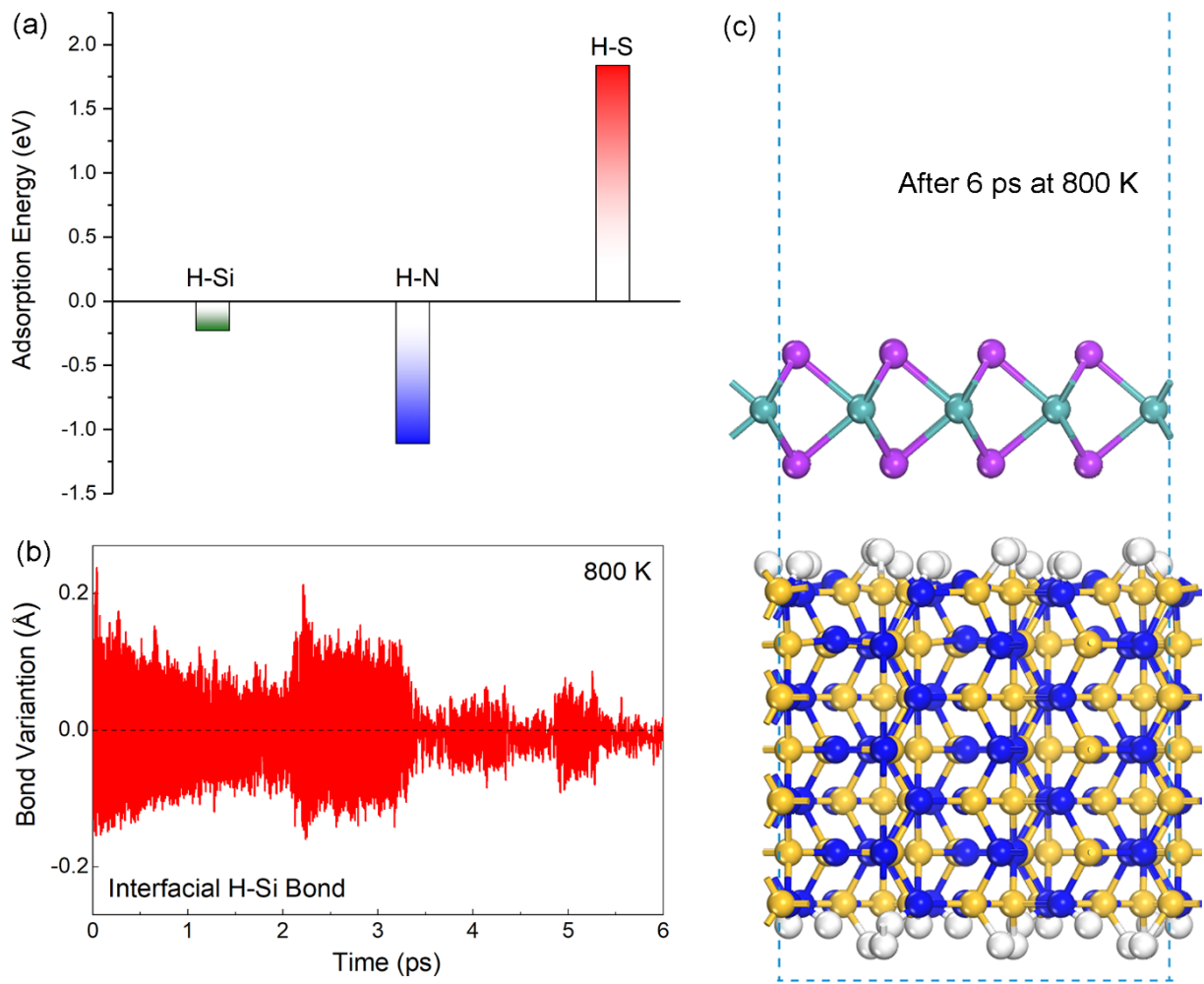

**Figure 3.** Hydrogenation selectivity and stability. (a) The hydrogen adsorption energy on *β*-$Si_3N_4$ (0001) and monolayer $MoS_2$. (b) The interface Si-H bond variation during the molecular dynamic (MD) simulation at the temperature of 800 K. (c) The final atomic structure (side view) of $MoS_2$ monolayer on the hydrogenated *β*-$Si_3N_4$ (0001) surface after the 6 ps MD simulation.

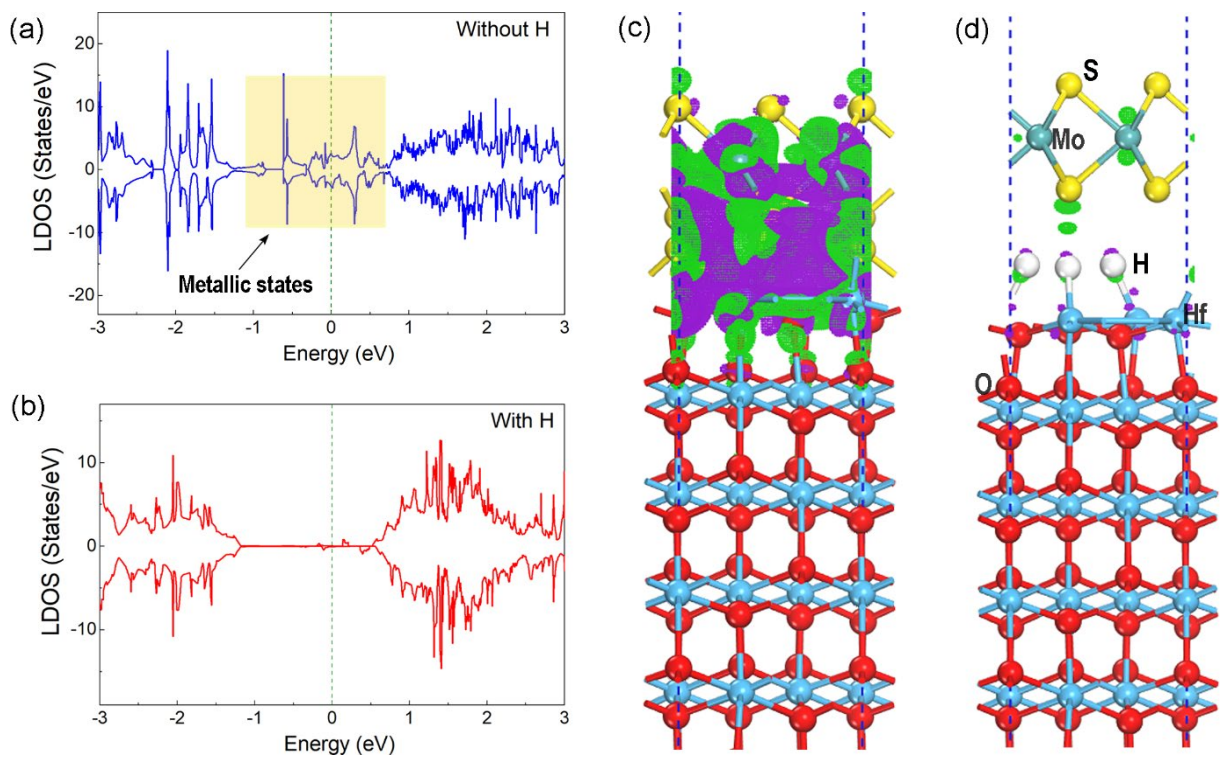

**Figure 4.** Hydrogenation effect on interface properties of monolayer MoS$_2$ on Hf-terminated HfO$_2$ (0001) surface. The LDOSs of MoS$_2$ monolayer on (a) HfO$_2$ (0001) surface and (b) hydrogen passivated HfO$_2$ (0001) surface. The charge density redistribution for MoS$_2$ monolayer on (ac) HfO$_2$ (0001) surface and (d) hydrogen passivated HfO$_2$ (0001) surface, where the green and purple dots denote the excess and accumulated charge density visualized by an iso-surface value of 3.0×10$^{-3}$ e/Å$^3$, respectively.